# Data Mining and Its Applications for Knowledge Management : A Literature Review from 2007 to 2012

Tipawan Silwattananusarn[1] and Assoc.Prof. Dr. KulthidaTuamsuk[2]

[1]Ph.D. Student in Information Studies Program,
Khon Kaen University, Thailand
[2]Head, Information & Communication Management Program,
Khon Kaen University, Thailand

**ABSTRACT:**

*Data mining is one of the most important steps of the knowledge discovery in databases process and is considered as significant subfield in knowledge management. Research in data mining continues growing in business and in learning organization over coming decades. This review paper explores the applications of data mining techniques which have been developed to support knowledge management process. The journal articles indexed in ScienceDirect Database from 2007 to 2012 are analyzed and classified. The discussion on the findings is divided into 4 topics: (i) knowledge resource; (ii) knowledge types and/or knowledge datasets; (iii) data mining tasks; and (iv) data mining techniques and applications used in knowledge management. The article first briefly describes the definition of data mining and data mining functionality. Then the knowledge management rationale and major knowledge management tools integrated in knowledge management cycle are described. Finally, the applications of data mining techniques in the process of knowledge management are summarized and discussed.*

## 1. INTRODUCTION

In information era, knowledge is becoming a crucial organizational resource that provides competitive advantage and giving rise to knowledge management (KM) initiatives. Many organizations have collected and stored vast amount of data. However, they are unable to discover valuable information hidden in the data by transforming these data into valuable and useful knowledge [2]. Managing knowledge resources can be a challenge. Many organizations are employing information technology in knowledge management to aid creation, sharing, integration, and distribution of knowledge.

Knowledge management is a process of data usage [6]. The basis of data mining is a process of using tools to extract useful knowledge from large datasets; data mining is an essential part of knowledge management [6].Wang & Wang (2008) point that data mining can be useful for KM in two main manners: (i) to share common knowledge of business intelligence (BI) context among data miners and (ii) to use data mining as a tool to extend human knowledge. Thus, data mining tools could help organizations to discover the hidden knowledge in the enormous amount of data.

As a part of data mining research, this paper focuses on surveying data mining applications in knowledge management through a literature review of articles from 2007 to 2012. The reason for reviewing research article this period is that data mining has emerged in KM research theme since 2006 [17] and it plays important roles as a link between business intelligence and knowledge management [26].

For article filtering, we search for the keyword "data mining" and "knowledge management" in the article title, abstract, and keywords fields on the Science Direct database.We limit the document search to date range published 2007 to 2012 and only document type of "research article" is included. The total number of documents published for this query by year shows in Figure 1.

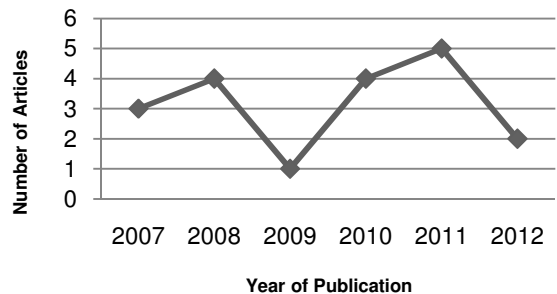

**Figure 1**  Number of Articles with "*Data Mining*" and "*Knowledge Management*" in the Title, Abstract, or Keyword Fields 2007-2012

The full text of each article is carefully reviewed to eliminate those articles that are not related to application of data mining in KM and not described how data mining could be employed or helped in KM. There are 10 articles related to these selection criteria.

Based on 10 articles on data mining applications for knowledge management, we survey and classify according to the six categories of data mining techniques: classification, regression, clustering, dependency modeling, deviation detection, and summarization.

The purpose of this paper is to review literature related to application of data mining techniques for KM in academic journals between 2007 and 2012. We organize this paper as follows: first, data mining definition and the data mining task primitively used in this study are described; second, the definition of knowledge management and the knowledge capture and creation tools are presented; third, articles about data mining in KM are analyzed and the results of the classification are reported; and last, the conclusions of the study are discussed.

# 2. DATA MINING

## 2.1 Definition of Data Mining

Data mining is an essential step in the knowledge discovery in databases (KDD) process that produces useful patterns or models from data (Figure 2) [7]. The terms of KDD and data mining are different. KDD refers to the overall process of discovering useful knowledge from data. Data mining refers to discover new patterns from a wealth of data in databases by focusing on the algorithms to extract useful knowledge [7].

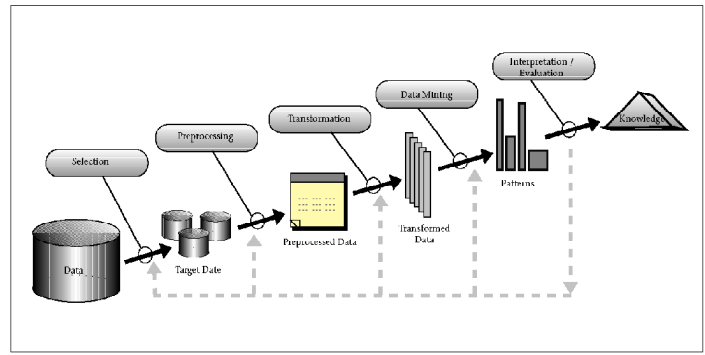

**Figure 2** Data Mining and the KDD Process(Source: Fayyad, et.al., 1996)

Based on figure 2, KDD process consists of iterative sequence methods as follows [7, 9]:

1. *Selection*: Selecting data relevant to the analysis task from the database

2. *Preprocessing*: Removing noise and inconsistent data; combining multiple data sources

3. *Transformation*: Transforming data into appropriate forms to perform data mining

4. *Data mining*: Choosing a data mining algorithm which is appropriate to pattern in the data; Extracting data patterns

5. *Interpretation/Evaluation* : Interpreting the patterns into knowledge by removing redundant or irrelevant patterns; Translating the useful patterns into terms that human-understandable

2.2 Data Mining Tasks

Fayyad et.al. (1996) define six main functions of data mining:

1. *Classification* is finding models that analyze and classify a data item into several predefined classes
2.
3. *Regression* is mapping a data item to a real-valued prediction variable

4. *Clustering* is identifying a finite set of categories or clusters to describe the data

5. *Dependency Modeling* (*Association Rule Learning*) is finding a model which describes significant dependencies between variables

6. *Deviation Detection* (*Anomaly Detection*) is discovering the most significant changes in the data

7. *Summarization* is finding a compact description for a subset of data

Data mining has two primary objectives of prediction and description. Prediction involves using some variables in data sets in order to predict unknown values of other relevant variables (*e.g. classification, regression, and anomaly detection*) Description involves finding human-understandable patterns and trends in the data (*e.g. clustering, association rule learning, and summarization*) [8].

## 3. KNOWLEDGE MANAGEMENT

### 3.1 Definition of Knowledge Management

There are various concepts of knowledge management. In this paper we use the definition of knowledge management by McInerney (2002):

"Knowledge management (KM) is an effort to increase useful knowledge within the organization. Ways to do this include encouraging communication, offering opportunities to learn, and promoting the sharing of appropriate knowledge artifacts"

This definition emphasizes the interaction aspect of knowledge management and organizational learning.

Knowledge management process focuses on knowledge flows and the process of creation, sharing, and distributing knowledge (Figure 3) [5]. Each of knowledge units of capture and creation, sharing and dissemination, and acquisition and application can be facilitated by information technology.

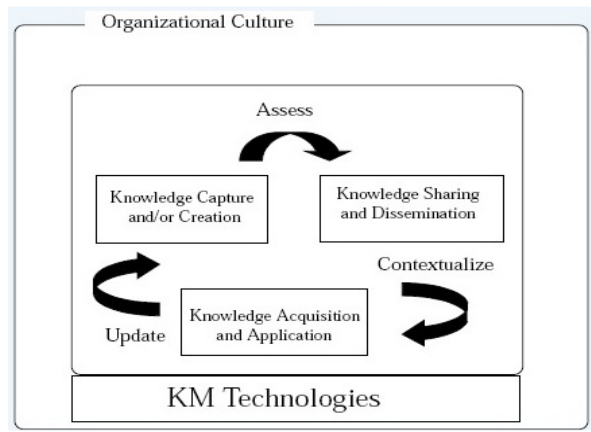

**Figure 3** KM Technologies Integrated KM Cycle (Source from Dalkir, K.,2005).

As technologies play an important role in KM, technologies stand to be a necessary tool for KM usage [1]. Thus, KM requires technologies to facilitate communication, collaboration, and content for better knowledge capture, sharing, dissemination, and application [5].

## 3.2 Knowledge Management: Capture and Creation Tools

This section provides an overview of a classification of KM technologies as tools and focuses on tools for capture and creation knowledge.

Liao (2003) classifies KM technologies using seven categories:

1. KM Framework
2. Knowledge-Based Systems (KBS)
3. Data Mining
4. Information and Communication Technology
5. Artificial Intelligence (AI)/Expert Systems (ES)
6. Database Technology (DT)
7. Modeling

Ruggles et.al. (1997) classify KM technologies as tools that generate knowledge (e.g. data mining), code knowledge, and transfer knowledge. Dalkir (2005) classifies KM tools according to the phase of the KM cycle (Figure 4). We can see that data mining involves in the part of knowledge creation and capture phase.

| Knowledge Creation and Capture Phase | Knowledge Sharing and Dissemination Phase | Knowledge Acquisition and Application Phase |
|---|---|---|
| Content creation | Communication and | E-learning technologies |
| ■ Authoring tools | collaboration technologies | ■ CBT |
| ■ Templates | ■ Telephone | ■ WBT |
| ■ Annotations | ■ Fax | ■ EPSS |
| ■ Data mining | ■ Videoconferencing | |
| ■ Expertise profiling | ■ Chat rooms | |
| ■ Blogs | ■ Instant messaging | |
| | ■ Internet telephony | |
| | ■ E-mail | |
| | ■ Discussion forums | |
| | ■ Groupware | |
| | ■ Wikis | |
| | ■ Workflow management | |
| Content management | Networking technologies | Artificial intelligence |
| ■ Metadata tagging | ■ Intranets | technologies |
| ■ Classification | ■ Extranets | ■ Expert systems |
| ■ Archiving | ■ Web servers, browsers | ■ DSS |
| ■ Personal KM | ■ Knowledge repository | ■ Customization– |
| | ■ Portal | personalization |
| | | ■ Push/pull technologies |
| | | ■ Recommender systems |
| | | ■ Visualization |
| | | ■ Knowledge maps |
| | | ■ Intelligent Agents |
| | | ■ Automated taxonomy systems |
| | | ■ Text analysis— summarization |

**Figure 4** Major KM Techniques, Tools, and Technologies (Source from Dalkir, K.,2005).

# 4. THE APPLICATIONS OF DATA MINING IN KNOWLEDGE MANAGEMENT

The reviews of ten articles has discussed on the applications of data mining to organizational knowledge management for effective capturing, storing and retrieving, and transferring knowledge. We divided the reviewed articles into four main groups: (i) knowledge resource; (ii) knowledge types and/or knowledge datasets; (iii) data mining tasks; and (iv) data mining techniques and applications used in KM. A detailed distribution of the ten articles categorized is shown in Table 1.

## 4.1 Knowledge Resources

In the study, we divided knowledge resources into eight groups as that which knowledge object to be stored and manipulated in KM and how data mining aids.

1. Health Care Organization: this domain was a use of the disease knowledge management system (KMS) of the hospital case study [10]. Data mining tool was used to explore diseases, operations, and tumors relationships. This tool used to build KMS to support clinical medicine in order to improve treatment quality [10].

2. Retailing: this was customer knowledge from household customers for product line and brand extension issues [14]; data mining can help and propose suggestions and solutions to the firm for product line and brand extensions. This doing by extracting market knowledge of customers, brands, products, and purchase data to fulfill the customers' demands behavior [14].

3. Financial/Banking: the domain knowledge covered financial and economic data; data mining can assist banking institutions making decision support and knowledge sharing processes to an enterprise bond classification [4].

4. Small and Middle Businesses (food company and food supply chain): there were two methods and processes to obtain knowledge resources: knowledge seeding-the relative knowledge to the problems; knowledge cultivating-the process to find the key knowledge from knowledge seeding [12]. Data mining and knowledge management integrated can help making better decisions [12]. As Death-On-Arrival (DOA) problem encountered in food supply chain networks (FSCN), Li et al. (2010) aimed to build Early Warning and Proactive Control (EW&PC) systems to solve such problems [13]. Knowledge Base was an important part of EW&PC systems. It contained data analysis by managers and organizes in an appropriate way for other managers. Data mining methods were helpful for the EW&PC systems [13].

5. Entrepreneurial Science:the knowledge resource was research assets in a knowledge institution [3]; there were three types of the research assets: research products, intellectual capital, and research programs. Data mining facilitated for knowledge extraction and helped guiding managers in determining strategies on knowledge-oriented organization competition [3].

**Table 1** Distribution of articles according to data mining and its applications

| Authors | Knowledge Resources | Knowledge Types | DM Tasks | DM techniques/ Applications |
|---|---|---|---|---|
| Lavrac et al. (2007) | Healthcare | Public Health Data <br>• *The health-care providers database* <br>• *The out-patient health-care statistics database* <br>• *The medical status database* | Classification; Clustering | Clustering Methods: <br>• *Agglomerative Classification;* <br>• *Principal Component Analysis;* <br>• *The Kolmogorov-Smirnov Test;* <br>• *The Quantile Range Test and PolarOrdination* <br>Classification – *C4.5* <br>**MediMap**: Visualization & Detection of Outliers |
| Hwang et al. (2008) | Healthcare (Clinical Diagnosis) | • Knowledge Conversion and Transfer <br>• Knowledge Measurement | Dependency Modeling | Data Mining Tool – IBM Intelligent Miner <br>Data Mining Techniques <br>• *Association Analysis* <br>• *Sequential Patterns Analysis* <br>Knowledge Management System (KMS) for Disease Classification |
| Liao, Chen & Wu (2008) | Retailing | Knowledge Extraction Customer Knowledge to product line and brand extension | Dependency Modeling; Clustering | *Apriori* Algorithms (Association Analysis) <br>*K*-means (Cluster Analysis) |
| Cheng, Lu &Sheu (2009) | Financial | • Knowledge Sets – *strings of data, models, parameters, and reports* <br>• Knowledge Sharing Processes to a Corporate Bond Classification | Classification Clustering | • Hybrid SOFM/LVQ Classifier for Bond Ratings <br>○ *The Self-Organizing Feature Map (SOFM)* <br>○ *The Learning Vector Quantization (LVQ)* <br>• Ontology of Knowledge Management and Knowledge Sharing <br>• Financial Knowledge Management System (FKMS) Prototype for Financial Research Purposes |
| Li, Zhu & Pan (2010) | Small & Middle Businesses (SMBs): Food Company | Knowledge Seeding &Knowledge Cultivating | Classification | • Extension Theory - *Extenics* <br>• Extension Data Mining (EDM) is combining *Extenics* with *Data Mining* |
| Li et al. (2010) | Food Supply Chain Networks | Knowledge Base | Classification | Decision Tree <br>Neural Network <br>Early Warning and Proactive Control Systems (EW&PC) |
| Cantu &Ceballos (2010) | Entrepreneurial Science | Research Assets | Classification | Data Mining Agents <br>• *Reasoning* <br>• *Pattern Recognition* <br>Knowledge-based System (KBS) <br>Knowledge and Information Network (KIN) Approach <br>Semantic Web Technologies |

| Wu et al. (2010) | Business | KM Styles & KM Performance | Classification | Bayesian Network Classifier Rough Set Theory |
|---|---|---|---|---|
| Liu & Lai (2011) | Collaboration and Teamwork Task (Worker's Log & Documents) | Knowledge Sharing Knowledge Graph Knowledge Flows (KFs) | Clustering | Process Mining Technique Knowledge Flow Mining Group-based Knowledge Flows (GKFs) |
| Ur-Rahman& Harding (2012) | Construction Industry | Textual Databases (Textual Data Formats) | Dependency Modeling; Clustering | Text mining<br>• *Clustering*<br>• *Apriori Association Rule Mining*<br>Multiple Key Term Phrasal Knowledge Sequences (MKTPKS) |

6. Business: data collected from questionnaire, an intensive literature review, and discussions with four KM experts [27]. Data mining can discover hidden patterns between KM and its performance for better KM implementations[27].

7. Collaboration and Teamwork: Worker's log and documents were analyzed each worker's referencing behavior and construct worker's knowledge flow. Data mining techniques can mine and construct group-based knowledge flows (GKFs) prototype for task-based groups [16].

8. Construction Industry: a large part of this enterprise information was available in the form of textual data formats [24]. This leads to the influence of text mining techniques to handle textual information source for industrial knowledge discovery and management solutions [24].

## 4.2 Knowledge Types

This section described knowledge types in 8 organization domains for data mining collaboration process in the knowledge creation.

- **Health-care System domain,** the dataset composed of three databases: the health-care providers' database; the out-patient health-care statistics database; and the medical status database [11].Another data source was from hospital inpatient medical records [10].

- **Construction Industry domain,** a sample data set was in the form of Post Project Reviews (PPRs) as defining good or bad information [24]. Multiple Key Term Phrasal Knowledge sequences (MKTPKS) formation was generated through applications of text mining and was used an essential part of the text analysis in the text documents classification[24].

- **Retailing domain:** customer data and the products purchased have been collected and stored in databases to mine whether the customers' purchase habits and behavior affect the product line and brand extensions or not [14].

- **Financial domain:** There were two datasets posed in financial domain: (i) to identify bond ratings, knowledge sets contained strings of data, models, parameters and reports for each analytical study; and (ii) to predict rating changes of bonds, cluster data of bond features as well as the model parameters were stored, classified, and applied to rating predictions [4].

- **Small and Middle Businesses (SMBs) domain:** Knowledge types in small and middle businesses in case of Food Company were related to the corporate conditions or goals of the problem among all departments to develop a decision system platform and then formed the knowledge tree to find relations by human-computer interaction method and optimize the process of decision making[12].To solve food supply chain networks problems, Li et al. (2010) developed EW&PC prototype which composed of major components of: (i) knowledge base, (ii) task classifier and template approaches, (iii) DM methods library with expert system for method selection, (iv) explorer and predictor, and (v) user interface [13]. This system built decision support models and helped managers to accomplish decision-making.

- **Research Assets domain:** In Cantu & Cellbos (2010) focused on managing knowledge assets by applied aknowledge and information network (KIN) approach. This platform contained three components types of research products, human resources or intellectual capital, and research programs. The various types of research assets were handled on domain ontologies and databases [3].

- **Business domain:** there were two types of knowledge attributes conducted: condition attributes and decision attribute [27]. Condition attributes included four independent attributes of the KM purpose, the explicit-oriented degree, the tacit-oriented degree, and the success factor. Decision attribute included one dependent attribute of the KM performance [27].

- **Collaboration and Teamwork domain:** a dataset used from a research laboratory in a research institute. It contained 14 knowledge workers, 424 research documents, and a workers' log as that recorded the time of document accessed and the documents of workers' needed [16]. For the workers' log, it was generated to 2 levels of codified-level knowledge flow and topic-level knowledge flow [16]. The two types of knowledge flow were determined to describe a worker's needs. To collect the knowledge flow, documents in the datasetwere categorized into eight clusters by data mining clustering approach [16].

## 4.3 Data Mining Techniques/Applications Used in Knowledge Management

Within the context of articles reviewed, applications of data mining have been widely used in various enterprises ranging from public health-care, construction industry, food company, retailing to finance. Each field can be supported by different data mining techniques which generally include classification, clustering, and dependency modeling. We provided a brief description of the four most used data mining techniques including its common tools used and some references as follows [7]:

**Classification:** Classification is one of the most common learning in data mining. This task aims at mapping a data item into one of several predefined classes. Examples of classification methods used as part of knowledge management include the classifying of the patients from primary health-care centers to specialists; the combination of the data mining and decision support approaches in planning of the regional health-care system; and the implementation of visualization method to facilitate KM and decision making processes [11]. In the financial company, Cheng, Lu & Sheu (2009) implemented an ontology-based approach of KM and knowledge sharing in financial knowledge management system (FKMS) and applied the hybrid SOFM/LVQ classifier of clustering and classification data mining techniques to classify corporate bonds [4]. For small and middle businesses: food company domain, data mining can improve decision-making by knowledge cultivating method namely *Extenics* and *Extension data mining (EDM)* [12]. This method was the integration of data mining and knowledge management, to develop a decision support system platform for better decisions [12]. To solve the death-on-arrival (DOA) in food supply chain networks, corporate manager selected variables that might have influence on DOA by using *"decision tree"* of data mining method; and used *"neural network"* to monitor potential DOA for prediction [13]. As knowledge assets played an important role in knowledge economies, Cantu & Ceballos (2010) employed data mining agents for extracting useful patterns to assist decision makers in generating benefits from the knowledge assets and used a knowledge information network (KIN) platform for managing the knowledge assets[3].In the business organizations with a large volume of works, such companies wanted to better understand what the hidden patterns between the KM and its performance, using the combination of data mining techniques: Bayesian Network (BN) classifier and Rough Set Theory (RST) in their business could help companies producing the KM to be performed effectively and achieve higher efficacy resulted [27]. Common tools used for classification are decision trees, neural network, Bayesian network and rough set theory.

**Clustering:** This involved seeking to identify a finite set of categories and grouping together objects that are similar to each other and dissimilar to the objects belonging to other clusters.This technique has been applied in many fields, for example:

- **Healthcare:** clustering categories and attributes used in analyzing the similarities between community health centers [11].
- **Retailing:** clustering the segmentation for possible product line and brand extension to identify market to customer clusters [14].
- **Financial/Banking:** identifying groups of corporate bond clusters according to the industry and a specific segment within an industry; then tuning cluster data for each industry as a template for predicting rating changes [4].
- **Construction Industry:** clustering textual data to discover groups of similar access patterns [24].
- **Collaboration and Teamwork:** identifying groups of workers with similar task-related information needs based on the similarities of workers' knowledge flow [16].

Common tools used for clustering include k-means, principal component analysis, the Kolmogorov-Smirnov test and the quantile range test and polar ordination.

**Dependency Modeling:** This concerned with finding a model that describes significant relationships between attribute sets. For example, it is widely used in healthcare to develop clinical pathway guidelines and provide an evidence-based medicine platform [10]. In medical records management, it is helpful for clinical decision making [10]. It could give better results in knowledge refinement through a use of this technique on the construction industry dataset [24]; this technique used to mine customer knowledge from household customers [14]. Common tools for dependency modeling are *Apriori* association rules and sequential pattern analysis.

As above, we can see that data mining techniques and applications in literature reveal different solutions to different KM problems in practice.

## 5. CONCLUSIONS

In organization, knowledge is an important resource. Management of knowledge resources has become a strong demand for development. Discovering the useful knowledge has also significant approach for management and decision making. As data mining is a main part of KM, this paper has identified ten articles related to data mining applications in KM, published between 2007 and 2012. This aims to give a research summary on the application of data mining in the KM technologies domain. The results presented in this paper have some assumptions:

- On the basis of the publication rates, research on the application of data mining in KM will increase in the future and cover the interest in different areas.
- The classification of data mining tasks is usually the employed model in organization for description and prediction. However, we will see the hybridization techniques e.g. association rule and clustering; classification and clustering etc. in order to solve different KM problems. This trend will give rising in the future.
- In the context of healthcare, one article used the visualization technique as a supplement to other data mining tasks. This visualization system could enhance and lead to better performance in decision making.
- KM is an interdisciplinary research area. Thus, in the future, KM development may need integration with different technologies and demand more methodologies to solve KM problems.
- KM applications development tends to support expert decision making and will be the application of a problem-oriented domain.

In this paper, we have shown that data mining can be integrated into KM framework and enhanced the KM process with better knowledge. It is clear that the data mining techniques will have a major impact on the practice of KM, and will present significance challenges for future knowledge and information systems research.